\documentclass[a4paper,11pt]{article}
\usepackage{pos}

\title{Non-Equilibrium Thermodynamics of Black-Hole Coronae: QPOs, Turbulence, and Jets}

\author*[a]{Vanessa L\'opez-Barquero}
\author[b,c]{Alejandro Jenkins}
\author[a]{Christopher S. Reynolds}
\author[d]{Andrew Fabian}

\affiliation[a]{Department of Astronomy, University of Maryland, College Park, MD 20742-2421, USA}

\affiliation[b]{Laboratorio de F\'isica Te\'orica y Computacional, Escuela de F\'isica, Universidad de Costa Rica, 11501-2060, San Jos\'e, Costa Rica}

\affiliation[c]{International Centre for Theory of Quantum Technologies (ICTQT), University of Gda\'nsk, 80-308, Gda\'nsk, Poland}

\affiliation[d] {Institute of Astronomy, University of Cambridge, Madingley Road, Cambridge CB3 OHA, UK}

\emailAdd{v.lopezbarquero@gmail.com}
\emailAdd{alejandro.jenkins@ucr.ac.cr}

\abstract{The variability of X-rays observed from accreting black hole systems, including quasi-periodic oscillations (QPOs), suggests a complex nonlinear dynamics in the corona.  Here, we propose a new theoretical framework for this variability, based on non-equilibrium thermodynamics.  In this model, coronal variability arises from feedback between a macroscopic oscillation of the plasma and the rate at which it is cooled by the inverse Compton scattering of soft photons from the disc.  The ``pair thermostat'' mechanism then allows the corona to act as a heat engine that extracts work cyclically from the underlying thermal disequilibrium between the low-entropy heating from the black hole and the high-entropy cooling by soft photons from the disk, in close analogy to the well-known $\kappa$-mechanism for pulsating stars.  This coronal self-oscillation may explain QPOs without invoking an external periodic driving.  Moreover, we argue that this mechanism can generate coronal turbulence and jets.}

\FullConference{
High Energy Phenomena in Relativistic Outflows IX (HEPRO-IX)\\
4-8 August 2025\\
Rio de Janeiro, Brazil\\}

\begin{document}
\maketitle

\section{Introduction}

Black hole (BH) environments constitute exceptional laboratories for probing extreme gravity and high-energy plasma physics. Central to many outstanding questions is the corona: a compact, optically thin region of hot electrons that upscatters lower-energy photons via Comptonization to produce the observed hard X-ray emission. Despite substantial progress enabled by modern spectral, timing, and polarimetric observations, the corona's geometry, structure, and physical processes powering its emission remain poorly constrained.  Since the corona plays a vital role in linking the accretion flow, magnetic fields, and radiation of these systems, addressing these questions is essential to progress in high-energy astrophysics.

A valuable diagnostic of coronal activity is its temporal variability, especially the so-called quasi-periodic oscillations (QPOs)~\cite{2019NewAR..8501524I}.  QPOs are manifest as narrow peaks in the X-ray power spectra of accreting BH systems and are widely interpreted as signatures of nearly periodic processes in the inner regions.  They are detected in the hard X-ray band, suggesting that their origin, or at least their expression, likely occurs in the corona. In X-ray binaries, low-frequency QPOs (LFQPOs) in the $\sim0.01$--$10~\mathrm{Hz}$ range show promise as probes of the physical conditions and fundamental parameters of the system in the immediate vicinity of the BH. However, despite extensive observational characterization and numerous theoretical proposals, their physical origin remains contested: proposed models range from Lense–Thirring precession to instabilities at the disk–corona interface and geometric modulation of the emitting region, but no single framework currently accounts for the full range of observed phenomena across spectral states.

The absence of comprehensive models presents challenges for using QPOs as system diagnostics. The link between QPO behavior, accretion states, spectral transitions, and jet ejections suggests that disk-based explanations may be insufficient. Instead, intrinsic variability in the corona may drive these phenomena.

This work outlines a new conceptual framework for understanding coronal variability, based on non-equilibrium thermodynamics and the theory of nonlinear dynamical systems. In this formulation, the corona operates as a self-oscillating system, which generates and maintains QPOs through an internal feedback mechanism.  Unlike forced or parametric resonances, a self-oscillator does not depend on matching the periodicity of an external power source and the oscillator's natural periodicity.  A self-oscillator is an autonomous {\it engine}, and the engine dynamics of the X-ray corona could be connected with active phenomena like the generation of turbulence and jets.

\section{Self-oscillators and engines}
\label{cycle}
A self-oscillator is an open system that generates and maintains a periodic or quasi-periodic variation at the expense of an external source of power that has no corresponding periodicity.\footnote{The same class of phenomena is also referred to in the literature by many other names, such as maintained, sustained, autonomous, self-induced, self-excited, or auto-oscillation.}  It is therefore qualitatively different from resonators, in which a significant amplitude is seen only when there is a precise matching of an external driving frequency to the natural frequency of the resonator.  Note that this is true even for non-linear or parametric resonators, where the matching need not be one-to-one, since in those cases the driving and the resonant frequency must be close to a low-order rational ratio \citep{1976mech.book.....L}.  One motivation for attempting to describe QPOs as self-oscillations is to obviate the need for an external periodicity capable of driving the QPO resonantly.

\begin{figure*}
\centering
    \includegraphics[width=0.5 \textwidth]{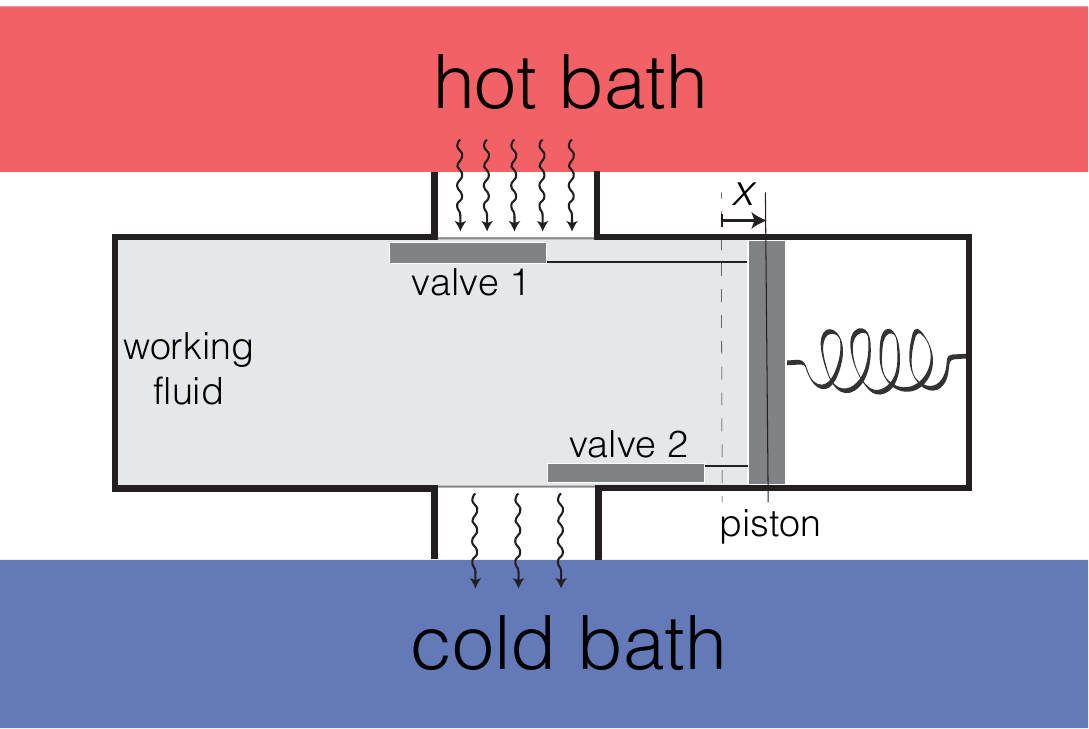}
    \caption{Diagram of a heat engine as a self-oscillator, powered by a feedback between the position $x$ of the piston and the coupling of the working fluid to two external baths at different temperatures.  As $x$ increases, the working fluid expands and valve 1 closes, reducing the rate of heating, while valve 2 opens, increasing the rate of cooling.  Conversely, as $x$ decreases, the working fluid contracts, valve 1 opens (increasing the rate of heating), and valve 2 closes (reducing the rate of cooling).  Image adapted from \cite{2017AnPhy.378...71A}.}
    \label{fig:Rayleigh}
\end{figure*}

In general, the physical conditions for a self-oscillator are:

\begin{enumerate}

    \item a thermodynamic disequilibrium, which allows a sustained flow of heat and/or matter to pass through the boundaries of an open system,

    \item a macroscopic, mechanical degree of freedom (analogous to the position $x$ of the piston in Fig.~\ref{fig:Rayleigh}) that is endowed with inertia and coupled to that open system, and

    \item a feedback between the mechanical degree of freedom and the coupling of the open system to the external disequilibrium, allowing for the cyclical extraction of work $W_{\rm net} > 0$ in accordance with the Rayleigh-Eddington criterion that we will cover in the next section.

\end{enumerate}

An important instance of an astrophysical self-oscillation is stellar pulsation~\citep{2015pust.book.....C}, in which the macroscopic oscillation of the volume of the stellar plasma is sustained against internal damping by a positive feedback between that oscillation and the rate at which the plasma is cooled by the escape of photons.  This is based on the so-called $\kappa$-mechanism: when the star is expanded, the lower temperature that results causes an outer layer of helium to relax to a singly ionized state, lowering the opacity and allowing more heat to escape from the star's interior.  When the star contracts, the increased temperature causes the helium layer to become doubly ionized, increasing its opacity and making the stellar interior more ``heat tight''.  This allows the pulsating star to act as an autonomous heat engine \cite{1963ARA&A...1..367Z, 2013PhR...525..167J}.  The frequency of the pulsation of the star is essentially the same as that of the free (adiabatic) acoustic oscillation of the plasma.  That frequency $f$ is controlled by the radius of the star $R$ and the speed of sound $v_s$ in the stellar plasma, according to
\begin{equation}
    f = c \, \frac{v_s}{R} ~,
    \label{eqn:f}
\end{equation}
where $c$ is a dimensionless number of order one.  As long as this frequency is not too fast compared to the rate at which the helium layer can change its macroscopic ionization state, the $\kappa$-mechanism is effective and can sustain the oscillation without the need to match any external driving frequency.

We will argue that the corona of an X-ray binary system offers a plausible analog to this $\kappa$-mechanism: The rate at which the plasma is cooled by the inverse Compton scattering of the incident soft photons from the disk depends on the optical depth of the plasma.  This optical depth, in turn, depends on the density of electron-positron pairs, which is mediated by the well-known ``pair thermostat'' effect~\cite{2017MNRAS.467.2566F,1985ApJ...289..514Z}. In this self-regulating mechanism, heating raises the plasma temperature until electron-positron pair production becomes efficient enough, whereupon the pair density increases as the temperature of the plasma {\it falls}, due to equipartition. This implies that the Compton cooling is most rapid when the corona is at a {\it lower} temperature. Conversely, pair annihilation causes the temperature to rise and the optical depth to fall, so that the corona will cool less effectively when its temperature is higher.  Thus, the corona is more ``heat tight'' when the temperature is highest.  Before we describe in more detail how this combination of Compton cooling and the pair thermostat may sustain a self-oscillation of the corona, let us first briefly review the general theory of thermal feedback in autonomous heat engines.

\subsection{Thermal feedback}

In the case of a homogeneous fluid, we can express the first law of thermodynamics as
\begin{equation}
\label{eqn:1stlaw}
    d U = - \delta W + \delta Q + \mu \, d {\cal N} ,
\end{equation}
where $U$ is the total internal energy, $Q$ the heat absorbed by the fluid, $W$ the work performed by it on its surroundings, $\cal N$ the quantity of matter, and $\mu$ its chemical potential.  We use the symbol $d$ for exact and $\delta$ for inexact differentials.  The net work that the fluid does over a cyclical transformation is therefore
\begin{equation}
\label{eqn:work}
    W_{\rm net} = \oint \delta W = \oint ( \delta Q + \mu \, d {\cal N} ) ,
\end{equation}
where we have used that $U$ must return to its initial value after a full cycle (i.e., $\oint dU = 0$).  Meanwhile, by the second law of thermodynamics (Clausius's theorem), the entropy generated by the cycle is
\begin{equation}
\label{eqn:Clausius}
\Sigma = - \oint \frac{\delta Q}{T} = - \oint \frac{\delta Q}{\bar T \left ( 1 + T_d / \bar T \right)} \geq 0 ~,
\end{equation}
where $T$ is the temperature of the source from which the working substance receives heat $\delta Q$ at each instant, $\bar T$ is the average temperature over the cycle's full period, and \hbox{$T_d \equiv T - \bar T$}.  Similarly, let $\mu$ be the instantaneous chemical potential at which matter enters the working substance and $\bar \mu$ its average over the cycle, so that \hbox{$\mu_d \equiv \mu - \bar \mu$}.  Combining (\ref{eqn:work}) and (\ref{eqn:Clausius}) we obtain:
\begin{equation}
\label{eqn:RE-full}
W_{\rm net} \leq \oint \delta Q \left( 1 - \frac{1}{1 + T_d / \bar T} \right) + \oint \mu \, d {\cal N} = \oint \frac{\delta Q \cdot T_d}{\bar T + T_d} + \oint \mu_d \, d {\cal N} ~.
\end{equation}
It is immediately clear that net work can be extracted cyclically only from an open system coupled to an external thermal or chemical disequilibrium, since otherwise $T_d = 0$ and $\mu_d = 0$ and (\ref{eqn:RE-full}) implies that $W_{\rm net} \leq 0$ for any cycle.  For a purely thermal engine, we take $\mu_d = 0$ and obtain that
\begin{equation}
    W_{\rm net} \leq \oint \frac{\delta Q \cdot T_d}{\bar T + T_d} \simeq \frac 1 {\bar T} \oint \delta Q \cdot T_d = \frac{1}{\bar T} \int_0^\tau T_d \, \dot Q \, dt~,
\label{eqn:RE}
\end{equation}
where $\delta Q = \dot Q \, dt$ and $\tau$ is the cycle's period.  For an active system (i.e., $W_{\rm net} > 0$) the heating rate $\dot Q$ must vary in phase with the temperature $T_d$ (strictly, the relative phase $\phi$ between them must be $-\pi/2 < \phi < \pi/2$, with $\phi = 0$ being the most favorable case).  We express this condition symbolically as
\begin{equation}
\dot Q \sim T_d ~.
\label{eqn:REs}
\end{equation}
(On the contrary, for passive systems $\dot Q \sim -T_d$.)  Eddington formulated the criterion of (\ref{eqn:REs}) in the context of the pulsation of variable stars.  The $\kappa$-mechanism mentioned above decreases the outflow of heat when the temperature is higher than average ($T_d > 0$), allowing the integral $\int_0^\tau T_d \, \dot Q \, dt$ in (\ref{eqn:RE}) to be positive.  For this reason, the $\kappa$-mechanism has also been called the ``Eddington valve''.

Eddington's use of (\ref{eqn:RE}) in the theory of stellar pulsation is equivalent to an earlier result by Rayleigh on the phase relation between the flow of heat and a thermoacoustic self-oscillation.  In mechanical engineering, this principle (that on average the air should lose heat more rapidly when expanded, and gain heat more rapidly when contracted, in order for the heat flow to encourage the acoustic oscillation) is known as the ``Rayleigh criterion''.  The general result of (\ref{eqn:RE-full}) has therefore been called the ``Rayleigh-Eddington criterion'' in recent work on the physical theory of self-oscillators and engines \cite{2017AnPhy.378...71A}.

\section{Compton cooling and pair thermostat in the corona}

\begin{figure*}
\centering
    \includegraphics[width=0.6 \textwidth]{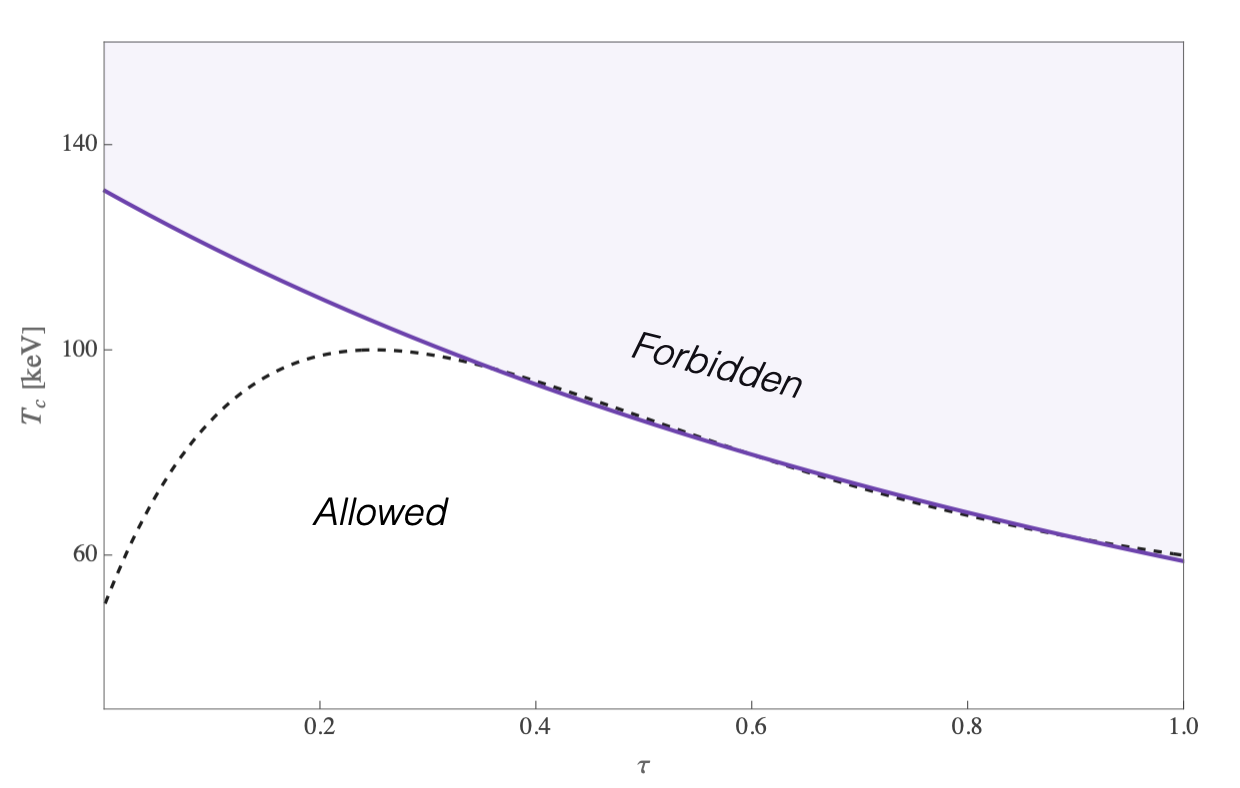}
    \caption{Schematic representation of the ``pair thermostat'' as a plot of the temperature $T_c$ of the corona versus its optical depth $\tau$.  The dotted line illustrates a possible transformation of the plasma as heat is injected.  Within the allowed region, more heat increases both $T_c$ and $\tau$.  When the plasma reaches the boundary of the forbidden region, further heating leads to an increase in the density of pairs (and therefore of $\tau$) while $T_c$ falls due to equipartition.  This image is adapted from \cite{1997ApJ...487..747D}.}
    \label{fig:PairThermostat}
\end{figure*}

The X-ray corona is principally cooled by inverse Compton scattering of soft photons from the disk.  The rate of this cooling depends on the optical depth of the plasma, and therefore on the density of pairs $n_p$, according to the phase relation $- \dot Q \sim n_p$.  If the coronal plasma is in the pair thermostat regime, then $T_d \sim -n_p$.  Thus, the combination of Compton cooling and pair thermostat implies that the corona fulfills the Rayleigh-Eddington criterion of (\ref{eqn:REs}) and can therefore act as an engine that extracts work cyclically from the underlying disequilibrium between the low-entropy heat injected into the corona and the high-entropy heat radiated into space.

Although the microscopic interactions within the plasma are very fast compared to the period of a macroscopic QPO, pair production prevents the plasma from approaching equilibrium during the heating phase.  This means that the entropy of the plasma is not a function of the macroscopic state variables and that the plasma does not obey an equation of state.  This is analogous to what happens to the working gas in an internal combustion engine: the gas does not approach equilibrium during combustion because many molecules collide with the moving piston, preventing the gas from achieving ``molecular chaos'' (Boltzmann's {\it Sto\ss zahlansatz}).
    
\section{A cycle for the corona}

\begin{figure*}
\centering
    \includegraphics[width=1 \textwidth]{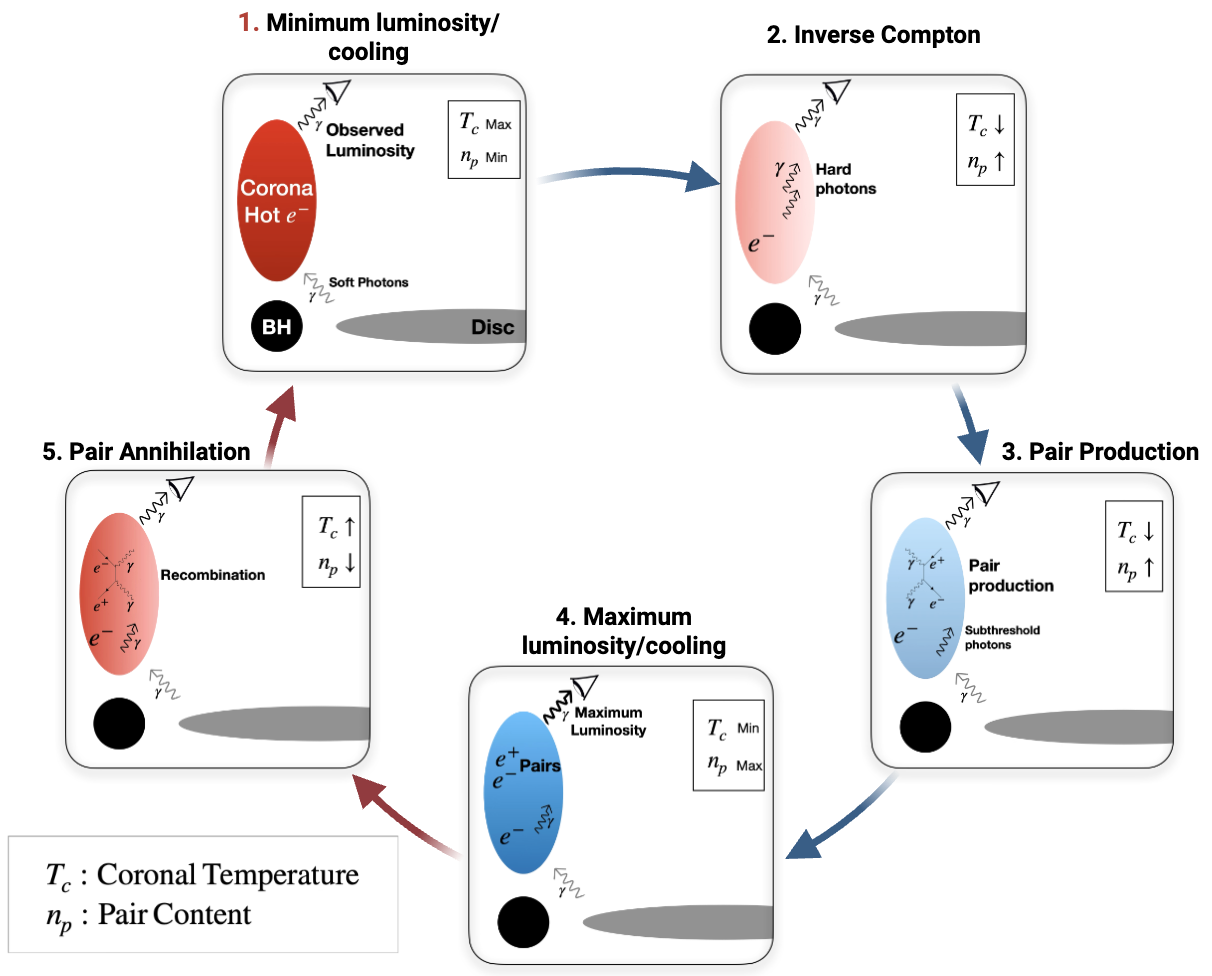}
    \caption{Proposed thermodynamic cycle of the corona, which would manifest itself as a QPO of the X-ray luminosity.  The work generated by this cycle may drive the generation of turbulence and jets in the corona.}
    \label{fig:cycle}
\end{figure*}

Based on the general considerations outlined above, we propose a thermodynamic cycle for the corona that can explain the QPO of the observed X-ray luminosity and which extracts work that sustains the macroscopic oscillation of the coronal plasma against its internal damping, while also potentially generating turbulence and jets.  This scheme is summarized in Figure \ref{fig:cycle}.

In step 1, the corona has reached its maximum temperature, and the density of electron-positron pairs is at its minimum.  The corona is constantly bombarded by soft photons from the nearby disk.  As those photons are up-scattered by the corona, the observed X-ray luminosity is generated.  In step 2, hard photons within the corona begin to produce electron-positron pairs.  In step 3, the temperature of the corona drops due to the ``pair thermostat'' effect, while the X-ray luminosity (and therefore the rate of cooling of the corona) increases with the increasing optical depth of the plasma.  In step 4, the corona reaches its lowest temperature and highest X-ray luminosity.  In step 5, the electron-positron pairs begin to annihilate, increasing the temperature of the plasma and reducing the observed luminosity.

The period of this cycle may be controlled by an acoustic or magnetoacoustic mode acting as the piston (e.g., we may suppose that $n_p \sim V$, where $V$ is the plasma's volume, which varies with the QPO's period).  The Raleigh-Eddington criterion (\ref{eqn:REs}) will be fulfilled, and the active cycle will therefore persist, as long as the QPO is slow [see, e.g., (\ref{eqn:f})] compared to the rate of pair production and annihilation in the coronal plasma.  (This is analogous to the condition that the combustion process be very fast relative to the period of the piston's motion in an internal combustion engine.)  From (\ref{eqn:RE}) we see that a small percent variation of the coronal temperature, $|T_d|/ \bar T$, can be compensated by a large heat input from the BH.

To obtain an equation of motion we may express the work output as $\delta W = {\cal F} \, dx$ for a generalized force $\cal F$ acting on the piston degree of freedom $x$.  We then express the first law of thermodynamics in the form
\begin{equation}
    W_{\rm net} = \oint {\cal F} dx = \oint \delta Q = (1 - g_\ast x) \, d\Pi ~,
\label{eq:force}
\end{equation}
where the function $\Pi$ describes the thermal feedback (analogous to the action of the valves in Fig.\ \ref{fig:Rayleigh}), while the factor $g_\ast$ is related to how $T_d$ varies with $x$.  The details of how such a dynamical description can be applied to the X-ray corona are left for future investigation.\footnote{Because the working substance remains far from equilibrium during the cycle, the entropy is not a state function and it is not appropriate to write $\delta Q = T \, dS$. Equation (\ref{eq:force}) is formulated in a manuscript by Carlos A.\ Guti\'errez, currently in preparation under the working title ``Equations of motion for engines with thermal feedback''.}

\section{Turbulence and jets}

If the work output of a self-oscillator went entirely into the motion of the piston $x$, then the amplitude of $x$ would grow without bound.  Linear damping cannot stabilize the amplitude.  In the kinematic van der Pol model, which is widely used in the mathematical literature on self-oscillation, the motion of $x$ is described by
\begin{equation}
\label{eqn:vdP}
    \ddot x - a \dot x + b x^2 \dot x + \omega^2 x = 0 \quad \hbox{for} \quad a,b > 0 ~,
\end{equation}
where the non-linear damping $b x^2 \dot x$ is needed to obtain a {\it limit cycle}. The anti-damping $-a \dot x$ comes from a feedback mechanism that is capable of overcoming the viscous damping present.  It can be shown that any initial condition, except the unstable fixed point $x(0) = 0$; $\dot x(0) = 0$ will asymptotically approach a closed curve in phase space that corresponds to a self-oscillation with amplitude $2 \sqrt{a/b}$; see, e.g., \cite{2013PhR...525..167J}.

The physical interpretation of this non-linear damping is that it represents the extraction of work from the self-oscillating piston.  In the case of the X-ray corona, there are two obvious mechanisms that may be associated with this work extraction: the generation of turbulence within the plasma, whose energy is eventually (but not directly) dissipated into internal heat, and the acceleration of material that escapes from the corona.  Such a mechanism provides a plausible route by which coronal power may be tapped to drive or energize jets that can be sustained as long as the corona remains in the pair-thermostat regime.  In other words, the Rayleigh-Eddington criterion of (\ref{eqn:REs}) allows the hard corona to run as a {\it jet engine}.

\section{Outlook}

We have offered a simple argument that the X-ray corona in a binary system can act as an autonomous heat engine, cyclically extracting work from the thermal disequilibrium between the high-temperature heat source associated with the BH and the low-temperature cooling by inverse Compton scattering of the soft photons from the disk.  Just as the piston in the engine of a car can move with greatly varying frequency, as long as its motion is not fast relative to the rate of combustion of the gasoline, various macroscopic modes of the X-ray corona (which may be acoustic or magnetoacoustic) can be maintained by the thermal feedback that we have described.  Since our proposal depends on the ``pair thermostat'', a clear prediction of this proposed model is that the QPOs and other related active phenomena should only be observed when the high-energy tail of the distribution for the photons has an appreciable component beyond the pair-production threshold of 1 MeV.

Naturally occurring heat engines are common in nature: we have mentioned pulsating stars; another heat engine familiar to astrophysicists are plasma convection cells.  Their dynamics in time, however, has not been much studied by theoretical physicists and the standard tools of equilibrium statistical mechanics are not applicable to them.  The work generated by the coronal engine that we have proposed could provide a major additional source of turbulence in the relevant magnetohydrodynamic models.  Moreover, it can explain the hints that QPOs and jets may be connected with the hard state of the corona.  We are currently working on a more detailed formulation of this dynamical model and a more precise comparison with the available data.

\section*{Acknowledgements}
We thank Carlos Guti\'errez for useful discussions. VLB thanks the European Research Council (ERC) for support under the European Union's Horizon 2020 research and innovation program (grant No. 834203) and the Department of Astronomy at the University of Maryland, College Park.  AJ was supported by the Vicerrectorate of Research of the University of Costa Rica, project no.\ 112-C1-716.


\bibliography{corona-so}{}
\bibliographystyle{aasjournal}

\end{document}